\documentclass[aps,twocolumn,floatfix,amsfonts,amsmath,amssymb,superscriptaddress,showpacs]{revtex4}
\usepackage{graphicx}

\usepackage[latin1]{inputenc}
\usepackage[T1]{fontenc}

\newcommand{\etal}{\textit{et al. }}
\newcommand{\KT}{k_{\rm B}T}

\newcommand{\rvec}{\mathbf{r}}

\begin{document}

    \title{Crystallization and flow induced by inhomogeneous activity}
    
    \author{Jaffar Hasnain}
    \affiliation{University of Vienna, Faculty of Physics, Boltzmanngasse 5, Vienna, Austria}
    \author{Georg Menzl}
    \affiliation{University of Vienna, Faculty of Physics, Boltzmanngasse 5, Vienna, Austria}
    \author{Swetlana Jungblut}
    \affiliation{University of Vienna, Faculty of Physics, Boltzmanngasse 5, Vienna, Austria}
    \author{Christoph Dellago}
    \affiliation{University of Vienna, Faculty of Physics, Boltzmanngasse 5, Vienna, Austria}

\begin{abstract}
Based upon recent experiments in which passive particles are made into active swimmers by illuminating them with laser light, we explore the effect of applying a light pattern on the sample, thereby creating activity inducing zones, or active patches. We simulate a system of interacting Brownian diffusers that become active swimmers when moving inside an active patch and analyze the structure and dynamics of the ensuing stationary state. We find that, in some respect, the impacts of an activity inhomogeneity qualitatively recemble those of a temperature gradient for regular patterns. For asymmetric patches, on the other hand, this analogy breaks down and we encounter stationary states specific for a partially active motion. 
\end{abstract}
\pacs{651498191513, corresponding author}
\maketitle

\section{Introduction}
The motion of bacteria, the action of molecular motors along proteins, and the collective motion of flocks are all part of the rapidly growing study of the nonequilibrium statistics of active matter \cite{ActiveMatterReview,ActiveMatterReview2,ActiveMatterReview3,ActiveMatterReview4}. Recently, a number of soft matter systems have been constructed to probe the collective behaviors of such systems \cite{ActivematterMechanisms}. This work is inspired by the family of experiments in which specially tailored micrometer sized particles can be made into active swimmers by illuminating them with laser light \cite{Palacci,BechingerActive}. In the experiments of Buttinoni \etal \cite{BechingerActive}, micrometer-sized silica beads were half-coated with a thin layer of graphite and suspended in a water-lutidine mixture that was kept close to the critical demixing concentration. By pouring this colloidal suspension in a cavity, the particles were confined to a quasi two-dimensional geometry and the sample was then illuminated by laser light. The wavelength of the laser was chosen so that only the carbon coated hemisphere of the silica beads was heated. At sufficiently high light intensity, the carbon half of the beads heated the surrounding solvent to such an extent that local demixing occurred. This created an asymmetric concentration gradient around the bead that induced directed motion perpendicular to the equator separating the hemispheres. Furthermore, it was shown that the light intensity was directly proportional to the swimming speed of the particles. 
\begin{figure}[t]
  \centering
  \includegraphics[width=0.95\columnwidth]{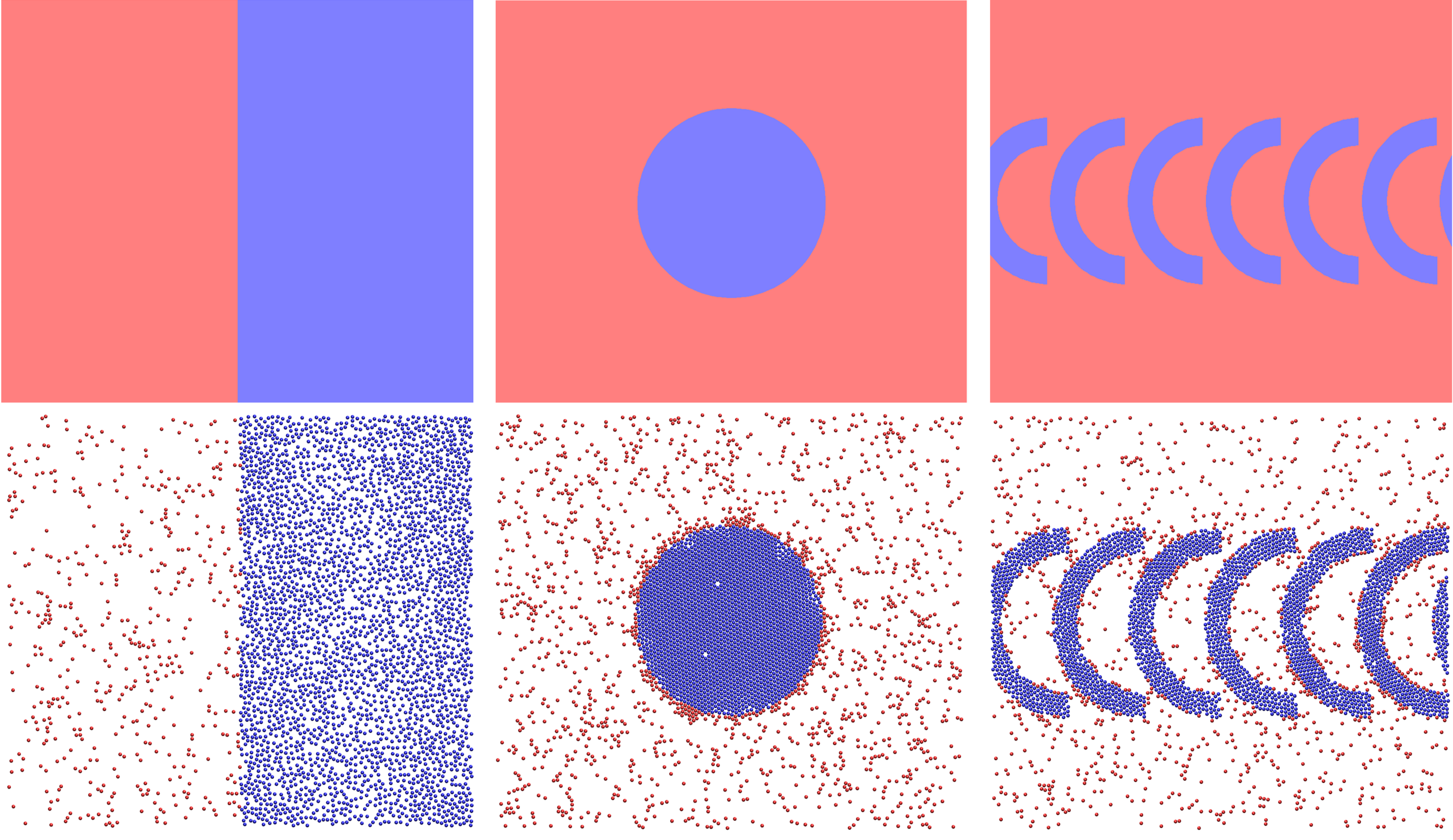}
  \caption{ The three different types of geometries considered in this work. The red regions correspond to areas that are illuminated by the activity-inducing laser light, and the particles in the blue region perform conventional Brownian diffusion. The bottom row depicts typical configurations of particles for large activity. The particles are colored according to whether or not they are active.}
  \label{Boundaries}
\end{figure}
The aim of this work is to study the effect of partially illuminating the sample, thereby creating active and inactive zones. **Proper introduction**

We shall consider the three different types of passive zones embedded in an active fluid, as depicted in the top row of Figure \ref{Boundaries} and analyze the ensuing nonequilibrium stationary state. 


\section{Model and methods}

Molecular dynamics simulations, discussed in Reference \cite{BechingerActive}, showed that although the colloidal particles are hard spheres, the Weeks--Chandler--Anderson (WCA) potential \cite{WCA} more accurately reproduces the radial distribution function obtained from experiment because the quasi two-dimensional arrangement of the particles allows for a small out-of-plane overlap. The potential energy between two particles interacting via the WCA potential is
\begin{eqnarray}
 U^{\mathrm{WCA}}(r)= \begin{cases} \epsilon \left[ (r_{0}/r)^{12}-2 (r_{0}/r)^{6} \right] + \epsilon &\mbox{if } r < r_0\\ 0 & \mbox{else,} \end{cases} 
\end{eqnarray}
where $r$ is the interparticle distance. The total potential energy of particle $i$ in the system is $U^{\textrm{tot}}_i=\sum_{i\neq j}U^{\mathrm{WCA}}(|\mathbf{r}_i-\mathbf{r}_j|) $ where $\mathbf{r}_i$ and $\mathbf{r}_j$ are the position vectors of the $i^\mathrm{th}$ and $j^\mathrm{th}$ particles in the system. The interaction strength between particles was set to $\epsilon=100\KT$, where $k_\mathrm{B}$ is the Boltzmann constant and $T$ is the temperature. The length scale of the interaction potential, $r_0$, is related to the Lennard-Jones parametrization by $r_0=2^{1/6}\sigma_{LJ}$ and represents the particle diameter. Experiments show that the particles obey overdamped Langevin dynamics without hydrodynamic interactions \cite{BechingerActive} and therefore the equation of motion for each particles is,
\begin{eqnarray}
 \gamma\dot{\mathbf{r}}_{i}&=&\mathbf{F}_{i}^{\textrm{WCA}}+\tilde{\mathbf{f}}_{i}+\gamma v(\mathbf{r}_i) \hat{\mathbf{e}}_i, \\
 \dot{\theta}_{i}&=&\tilde{\mathbf{\Gamma}}_{i}. 
 \label{Eqnmotion}
\end{eqnarray}
Here, $\dot{\mathbf{r}}_i$ is the velocity of particle $i$ and $\gamma$ is the friction constant related to the diffusion constant $D_0$ of a single particle in the fluid by the Einstein relation $\gamma=\KT/D_0$. The interparticle force, $\mathbf{F}_{i}^{\textrm{WCA}}$, is the negative gradient of the potential $U^{\textrm{tot}}_i$. The particles experience a random stochastic force $\tilde{\mathbf{f}}_{i}$ due to the solvent which was simulated by a Gaussian distribution that is delta correlated in time, and has a variance of $2 D_0 \gamma^2$. The last term, $\gamma v(\mathbf{r}_i) \hat{\mathbf{e}}_i$, is responsible for the active motion of the particle. This force acts along the orientation $\hat{\mathbf{e}}_i$ of the particle with position dependent magnitude $v(\mathbf{r}_i)$. In this work, the magnitude of the active force, $v(\mathbf{r}_i)$, is either a constant $v$ or $0$, depending on whether the particle is in the illuminated (red) or passive (blue) region in Figure \ref{Boundaries}. The orientation vector of each particle, $\hat{\mathbf{e}}_i = \{\cos \theta, \sin \theta \}$, performs an independent random walk in $\theta$, such that $\langle \theta(t) \theta(0) \rangle= 2 D_r \delta(t)$ and the rotational diffusion constant of the particles obeys the no slip relation, $D_r = 3 D_0 /r_{0}^2$.

In all of our simulations, we set $\KT$, $\gamma$, and $r_0$ to unity. As a result the diffusion constant, $D_0$, is also unity. We considered a system of $3600$ particles with a global (\textit{i.e.} across both zones) number density of $\rho_g=0.2886$, similar to experiment \cite{BechingerActive}. The simulation cell had width $L_x=120 r_0$ and height $L_y=L_x \sqrt{3}/2$. The system was subject to periodic boundary conditions and the simulations were initialized by arranging the particles in a hexagonal lattice that melts quickly due to the fact that the packing fraction of the system is far from crystallization. The systems were equilibrated for $5 \times 10^6$ timesteps and data were then gathered from trajectories of length $1.25 \times 10^7$. The phase behavior of a two-dimensional system with WCA interactions can approximated accurately by taking advantage of the fact that, for sufficiently high interaction strengths, the WCA potential can be mapped onto a hard sphere system with diameter $\sigma_{\textrm{HS}}^2=2 B_2/\pi$ where $B_2$ is the second Virial coefficient of the WCA potential \cite{HSEOS}. For $\epsilon=100\KT$, $\sigma_{\textrm{HS}}^2= 0.9861 r_0$ and therefore, the hard sphere packing fraction corresponding to $rho_g$ is $\eta_g=  \rho_g \pi \sigma_{\textrm{HS}}^2/4 = 0.2205$. Typically, the activity of a particle is quantified by its P{\'e}clet number, $\mathrm{Pe}=r_0 v/D_0$ and in experiments values of about $\mathrm{Pe}=200$ were achieved. The highest activities used in this study correspond to $\mathrm{Pe}=150$ and are within experimentally accessible range \cite{BechingerActive}. The passive patches that were considered consist of a halfplane of width $L_x/2$, a circle with radius $24\, r_0$, and six semicircular stripes of width $6\,r_0$ and outer radius $24\,r_0$. 

\section{Results and discussion}

\subsection{Comparison between active and diffusive patches}

In order to study the effect of the activity beyond a simple increase in the mean square displacement (MSD), we also simulated a system of Brownian particles with position dependent diffusion constant. Conceptually, this corresponds to a setup in which Brownian particles are subjected to a temperature difference. To compare the system with an activity difference to the system with a temperature difference, consider the MSD of a solitary active particle as reported in Refs. \cite{GolestanianActive,BechingerActive},
\begin{equation}
 \langle \Delta \mathbf{r}^2 \rangle = 4 D_0 t + \frac{2 v^2}{D_{r}^2}\left ( e^{- D_r t} + D_r t - 1 \right ).
\end{equation}
For short times, this MSD can be approximated by $\langle \Delta \mathbf{r}^2 \rangle = 4 D_0 t + (vt)^2 $, which is independent of the rotational diffusion constant. On the other hand, in the long time limit the MSD converges to $\langle \Delta \mathbf{r}^2 \rangle = 4 \left [D_0 + v^2/(2 D_r) \right ]t - v^2 /D_{r}^2$. One can therefore associate, for different levels of activity $v$, a long time diffusion constant $D_A=D_0 + v^2/2D_r$, which is the slope of the mean squared displacement for large $t$. In the following, the activity of the particles will be given in terms of $D_A$. 

Before we proceed, consider the even simpler reference model of an ideal gas of either active particles or Brownian diffusers whose mobility is position dependent using the half plane geometry in the first column of Figure \ref{Boundaries}. Since the particles in the red region are more mobile than those in the blue region, there is a net flow of particles from the more mobile region to the passive region. The system arrives at a stationary state when the density difference has compensated the imposed mobility difference. The stationary state of this system obeys the Smoluchowski equation \cite{FPEActiveparticles},
\begin{equation}
 \Delta \left [ D(\rvec)P(\rvec,\theta) \right]  + D_r \frac{\partial^2 P(\rvec,\theta) }{\partial \theta^2} - \hat{\mathbf{e}} \mathbf{\nabla} \left[ v(\rvec)P(\rvec,\theta) \right] = 0,
\end{equation}
where $\Delta$ and $\mathbf{\nabla}$ are the Laplace and gradient operators, respectively. If one imposes a difference in diffusivity with no activity, $v(\rvec)=0$, then the equation decouples in position and orientation angle and one obtains the condition $P(\rvec)D(\rvec)=c$ where c is an integration constant. For a step-like diffusion field $D(\rvec)$, one therefore obtains a step-like density profile, and the ratio, $\rho_0/\rho_H$ of the densities grows linearly with the ratio, $D_H/D_0$, of the diffusion constants. For the active patch system, $D(\rvec)=D_0$, and $v(\rvec)$ is a step function that is $v$ in the left half of the simulation cell and $0$ in the right half. 

The Smoluchowski equation for an ideal gas of active particles was solved approximately by Bialk\'e \etal \cite{FPEActiveparticlesEPL} and predicts that the ratio of the densities in the passive and active zones, $\rho_0/\rho_A$, scales with the square root of the ratio of the long time diffusion constants $D_A/D_0$. 

\begin{figure}[t]
  \centering
  \includegraphics[width=0.95\columnwidth]{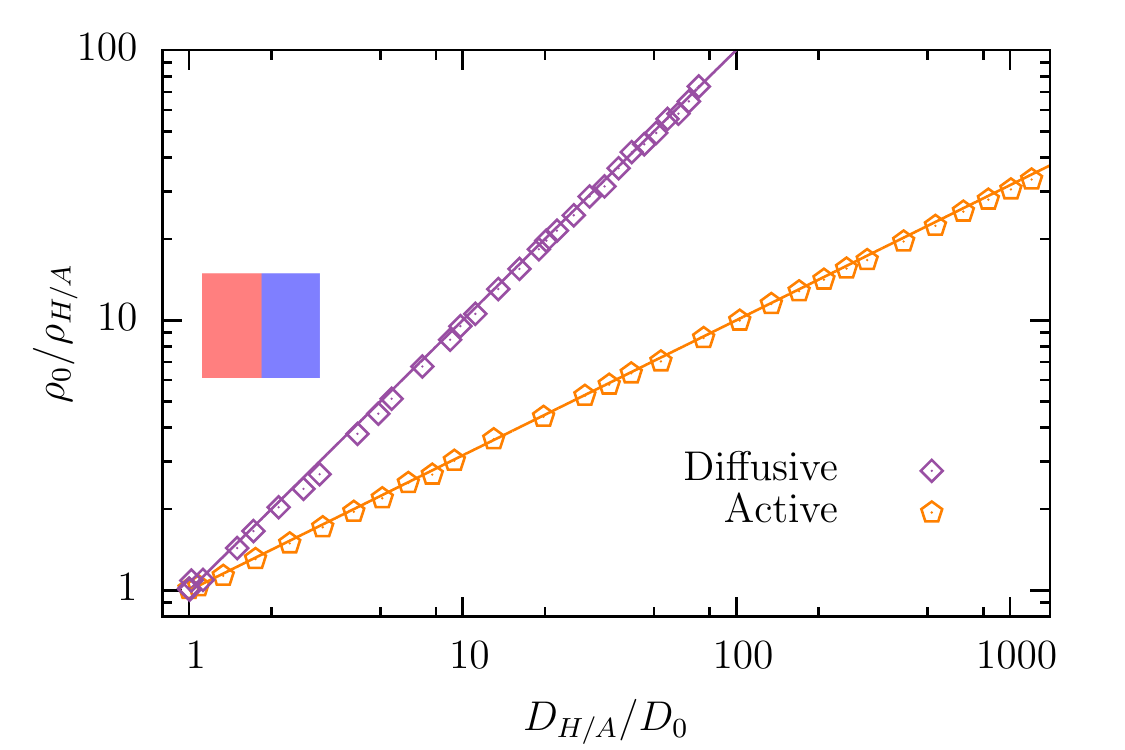}
  \caption{ Ratio of the densities of an ideal gas of particles for a half-plane geometry as a function of an imposed mobility difference that is either due to a larger diffusivity, $D_H/D_0$, (diamonds) or activity, $D_A/D_0$ (pentagons).}
  \label{IdealGPic}
\end{figure}

In Figure \ref{IdealGPic}, we plotted the ratio of the densities, $\rho_0/\rho_{H/A}$, in the blue and red regions as a function of the ratio of the respective diffusion constants, $D_{H/A}/D_0$. The purple diamonds represent the data obtained from simulations of a diffusivity difference, the orange pentagons were obtained from a system with an activity difference, and the lines are the theoretical predictions of the Smoluchowski equation that agree well with the simulation data.

\begin{figure}[t]
    \centering
     \includegraphics[width=0.95\columnwidth]{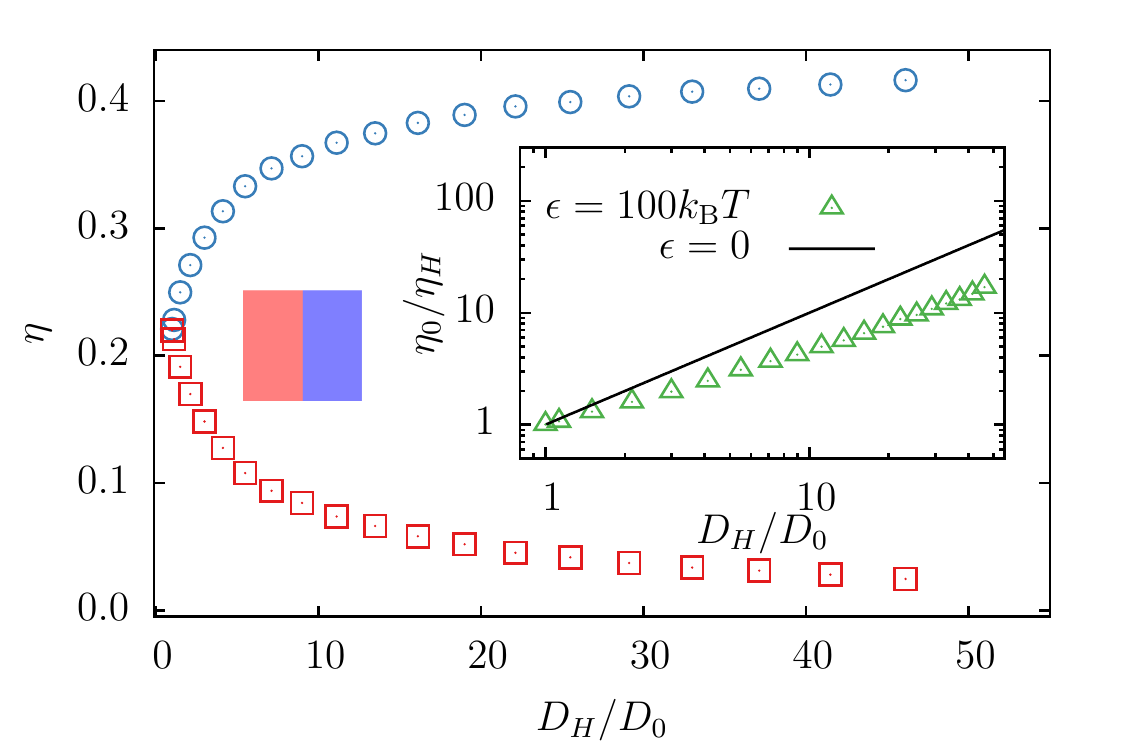}
     \includegraphics[width=0.95\columnwidth]{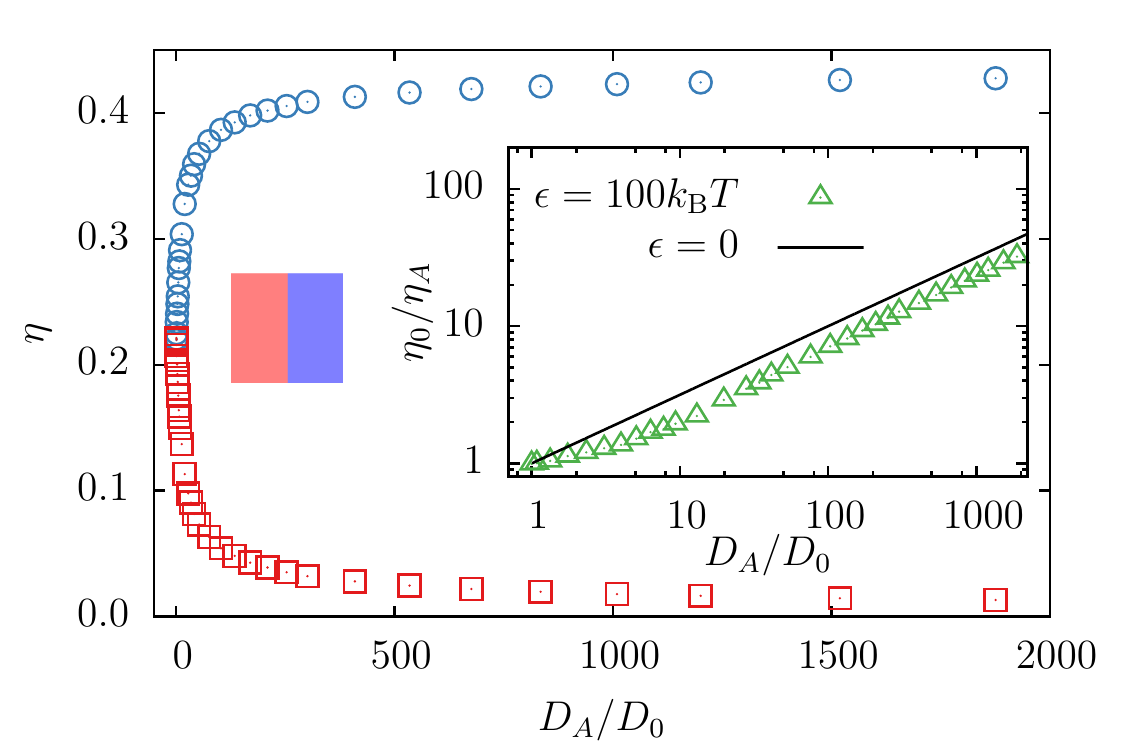}
    \caption{Packing fraction $\eta$ for each half-plane as a function of a temperature difference $D_H/D_0$ (top) and activity difference $D_A/D_0$ (bottom). For constant particle number, $\eta_{\mathrm{red}} A_{\mathrm{red}}+ \eta_{\mathrm{blue}} A_{{\mathrm{blue}}}= \eta_g A_g$, where $A_{\mathrm{red}}$, $A_{\mathrm{blue}}$, and $A_{g}$ are the areas of the red, blue and entire system, respectively. \textit{Insets:} Ratio of the densities is plotted on a double logarithmic scale and the black lines represent the ideal gas solutions.}
  \label{Densities}
\end{figure}

Having established the behavior of the ideal gas, we now consider the same patch for particles with interparticle interactions. Henceforth, in order to account for the excluded volume due to the interparticle interactions, the density differences will be measured in terms of packing fractions $\eta=\rho \pi \sigma^{2}_{\textrm{HS}}/4$. In Figure \ref{Densities}, we have plotted the packing fraction in each zone as a function of the mobility difference for the half plane geometry. In the inset, the ratio of the packing fractions is drawn on a double logarithmic plot and the black lines correspond to the ideal gas solution. Evidently, the interparticle interactions reduce the efficacy with which a mobility difference induces a density difference. In the supplemental video V1, the dynamics of a system with a diffusivity difference is compared to the dynamics of a system with an activity difference. The values of $D_H$ and $D_A$ were chosen such that the density difference for the two systems is the same. Both in the video and in the bottom left panel of Figure \ref{Boundaries}, one can see that the active particles form characteristic dynamical clusters for high levels of activity \cite{ActiveCrystallization,HaganActiveMat}. 

 \begin{figure}[t]
    \centering
      \includegraphics[width=0.95\columnwidth]{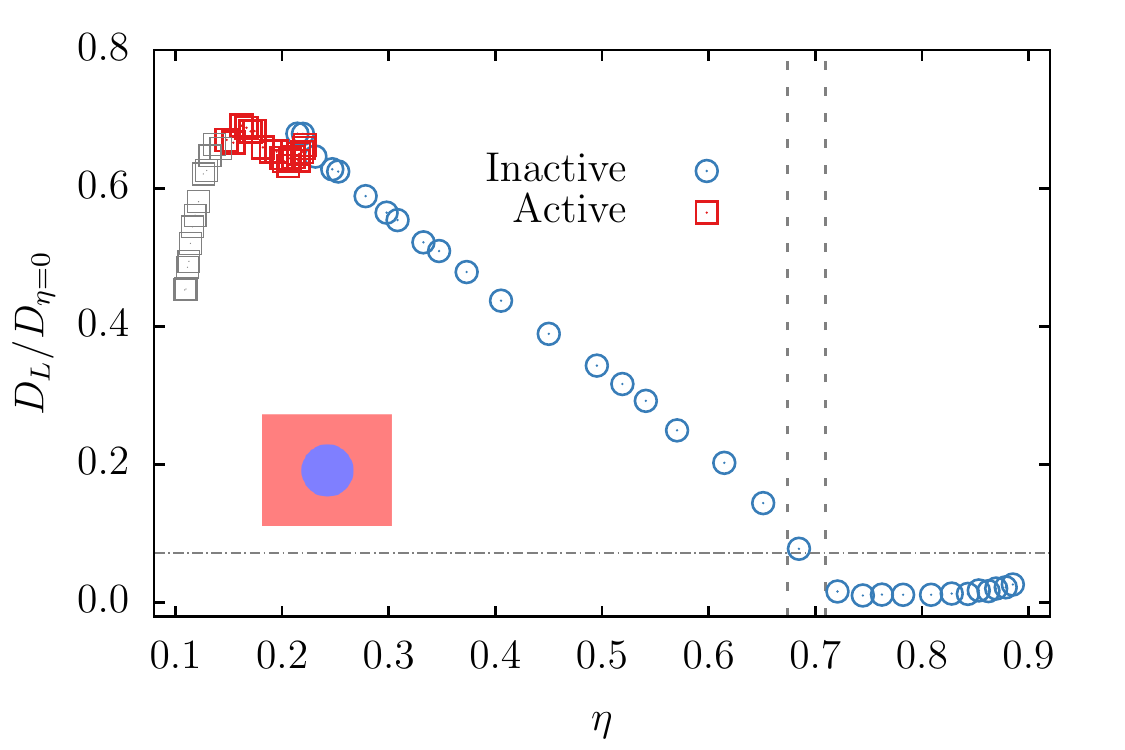}
    \caption{Effective diffusion constant of the particles in the active and inactive zones as a function of the packing fraction. The vertical gray lines are the freezing and melting packing fractions for hard disks and the horizontal gray line is the value of the effective diffusion constant at the freezing transition. The data points in gray are the effective diffusion constants of the active particles after crystallization has taken place in the passive zone.}
    \label{Circle}
  \end{figure}
\subsection{Crystallization in the passive patch and dynamical clustering at its boundary}

\begin{figure}[t]
    \centering
      \includegraphics[width=0.95\columnwidth]{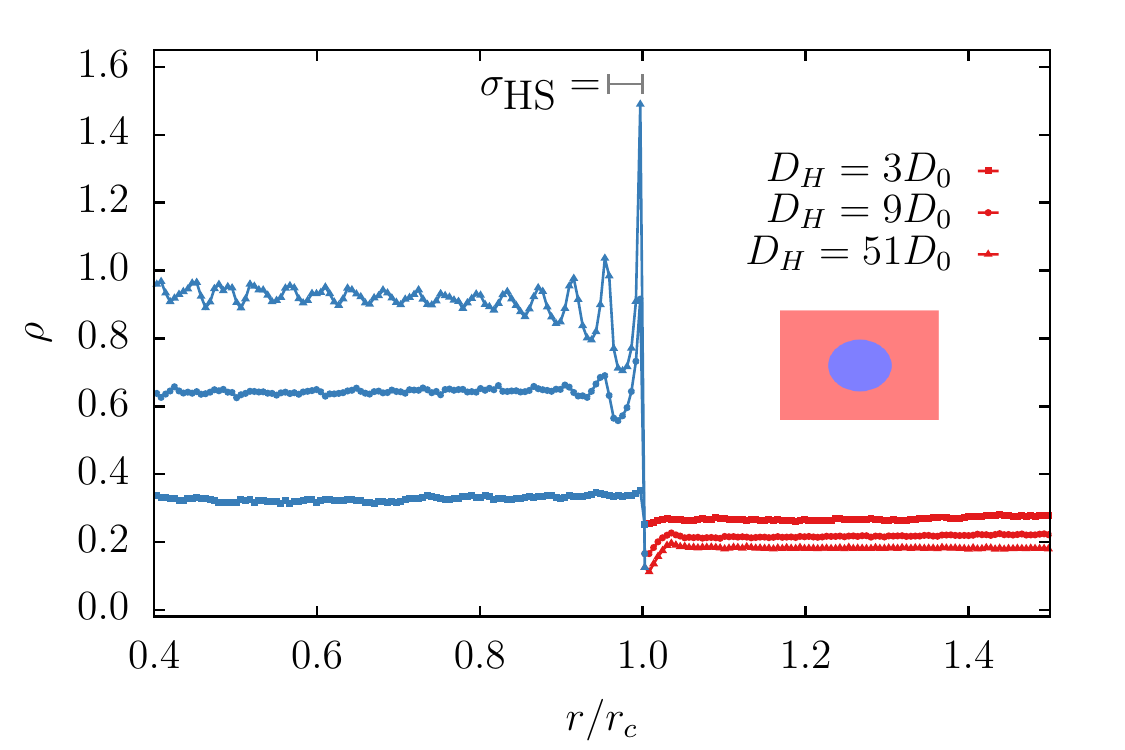}
      \includegraphics[width=0.95\columnwidth]{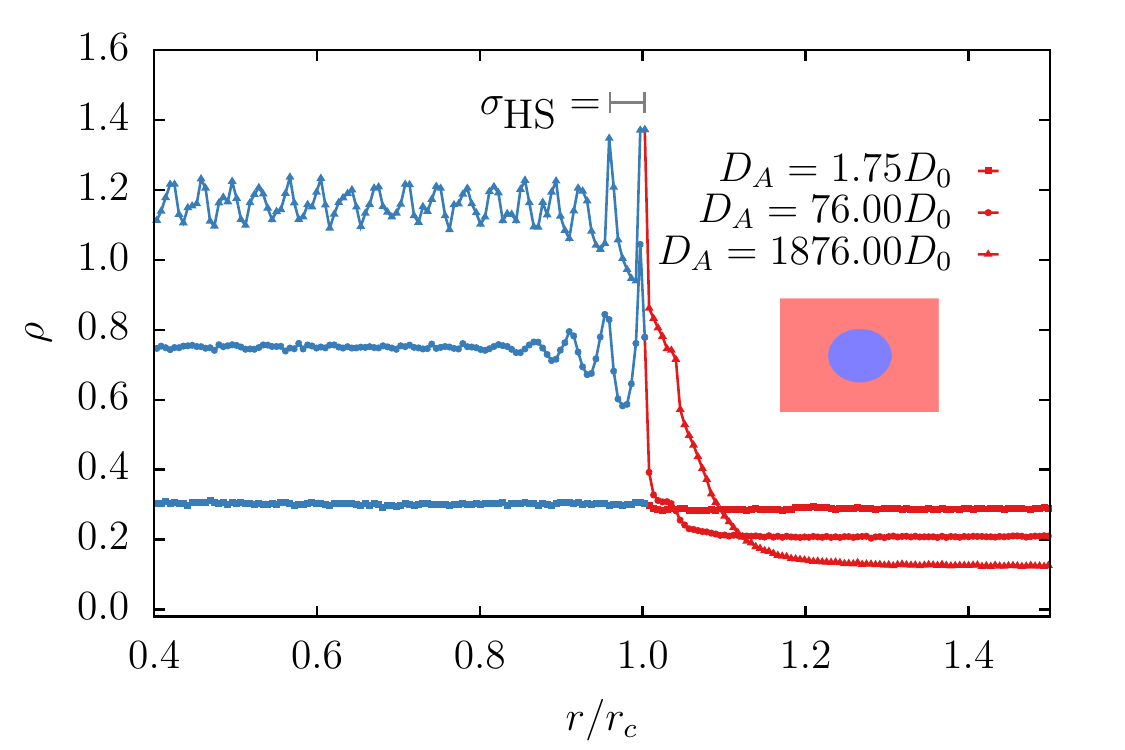}
    \caption{Local particle density, $\rho(r)$, as a function of the distance from the center of the circular passive zone, $r$, for low, intermediate, and high mobility differences. The $x$ axis has been rescaled by the radius of the passive zone $r_c=24\, r_0$ and the data points in blue correspond to the density inside the passive circle and the red data points represent the density outside of the passive zone. The effective hard sphere diameter $\sigma_{\textrm{HS}}$ has been drawn as a guide to the eye.}
    \label{DensityProfile}
  \end{figure}

In the case of the half plane, $\eta_g$ was such that in the limit of very large diffusion differences the density in the blue zone approaches $2\eta_g$, which is below the crystallization packing fraction of hard disks. Next we consider a passive circle in an active bath (middle column of Figure \ref{Boundaries}), that was small enough, so that for large differences in activity, the packing fraction in the passive zone was large enough to induce crystallization. As the density in the passive zone increases due to rising activity differences, the mobility of the particles in the passive zone decreases until a freezing transition takes place, after which the particles oscillate about their lattice sites.In Figure \ref{Circle}, we plotted the effective diffusion constants of the particles in the active (red squares) and passive (blue circles) regions as a function of the local packing fraction. The long time diffusion constants were obtained by dividing our trajectories into short segments in which no particles cross from one zone to another and then computing the mean square displacements of the particles in each zone separately. By rescaling the long time diffusion constants of the particles by the diffusion constant in absence of collisions, $D_{L} / D_{\eta=0}$, we obtained the effective diffusion constant plotted on the $y$-axis in Figure \ref{Circle}. For a dilute system, this ratio is unity by construction. Two independent criteria were used to determine whether crystallization took place and are indicated by the horizontal and vertical lines in the plot. The vertical gray lines represent the packing fraction at which the freezing (left vertical line, $\eta=0.674$) and the melting transition takes place (right vertical line $\eta=0.71$) in a hard disk system \cite{HSFreezing} and the density region between these two points corresponds to a fluid--solid coexistence regime. The second criterion used to identify the freezing transition is due to L{\"o}wen \cite{FreezingCriterion2D}, who found that the effective diffusion constant of hard disks has a value of $D_{L}/D_{\eta=0}=0.072$ (horizontal gray line) along the melting line. The data shown in Figure \ref{Circle} are consistent with a freezing transition that is corroborated by visual inspection of trajectories of the system (see Supplemental Video $2$). For the particles in the active zone, the measurement of the effective diffusion constant is strongly affected by the presence of the passive zone. For small activity differences, the effective diffusion constant in the red zone increases because the density in that zone decreases. However, for large activity differences, after crystallization has taken place, the active particles travel so quickly that a collision with the boundary of the passive zone is very likely. After such a collision has taken place, the active particles continue to swim in the same direction and persists at the boundary of the crystalline domain. The active particles therefore aggregate at the boundary of the crystalline zone when the swimming speed is large. This has the effect of reducing the mean square displacement of the particles and therefore one observes a reduction in the effective diffusion constant (gray squares in Figure \ref{Circle}) of the particles in the active zone. In the supplemental video V2 and in the snapshot of the configuration in Figure \ref{Boundaries}, one can observe the aggregation of active particles at the boundary of the crystalline domain that leads to this reduction in the effective diffusion constant. To illustrate the aggregation of the active particles at the boundary of the passive zone we plotted the density of the particles as a function of the distance from the center of the circle both for the diffusivity (top) and activity (bottom) induced crystallization in Figure \ref{DensityProfile}. For small mobility differences the density in each region is essentially homogenous. For intermediate mobility differences, there is a depletion zone of passive particles (blue data points) that has a width of one particle diameter $\sigma_{\textrm{HS}}$ that is due to the interactions between particles close to the boundary (we did not observe similar depletion zones in simulations of an ideal gas). For mobility differences crystallization takes place in both cases and the regularly spaced peaks in the density profile of the passive zone reflects the periodic are one particle diameter in width and are a signature of the crystalline domain. The biggest difference between the diffusive and the active case is seen in the density profile outside of the passive circle, after crystallization has taken place. The shoulder in the density profile in the bottom of Figure \ref{DensityProfile} for $D_A=1876\,D_0$ is two particle diameters wide and is a clear indication of the aggregation of active particles at the boundary. The active particles that aggregate at the boundary of the crystalline domain shear and rotate it stochastically, reminiscent of the dynamical clusters formed by bacteria \cite{PoonRotation}.
Finally, since the circular domain is incommensurate with a hexagonal crystal, one regularly observes the appearance of defect and dislocations that travel through the domain and they appear more frequently in the density range of the fluid-solid coexistence. This type of activity difference induced crystallization can easily be achieved experimentally, providing an easy opportunity to study defect and dislocation migration in arbitrarily shaped, hexagonally ordered, domains. Similar simulations and results have been recently published by Magiera \etal \cite{ActiveCircle}.


 \begin{figure}[t]
    \centering
      \includegraphics[width=0.95\columnwidth]{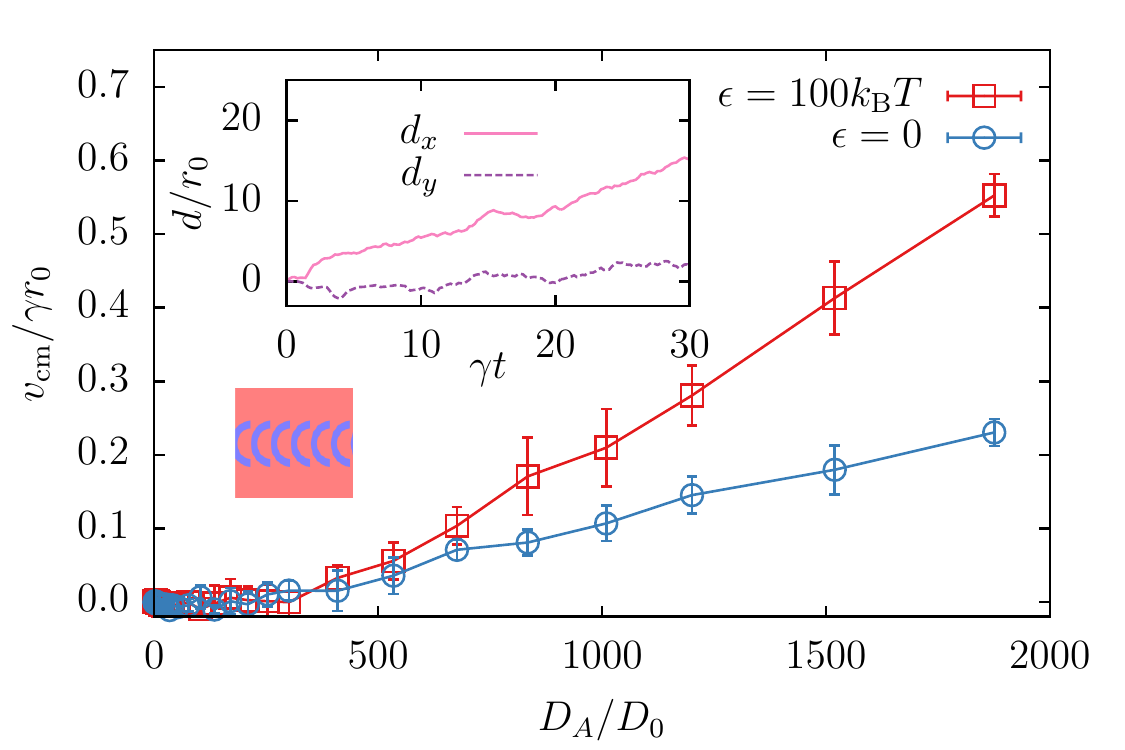}
    \caption{Mean velocity of the center of mass of the system as a function of the activity difference. Inset: mean displacement of the center of mass of the hard sphere system in the $x$- and $y$-direction for $D_A=1876 D_0$, as a function of time.}
    \label{ParticleCannons}
  \end{figure}

\subsection{Asymmetric passive patches}

The final active patch that we will consider consists of six semicircular passive stripes embedded in an active fluid, as shown in the third column of Figure \ref{Boundaries}. The significant difference between this type of patch and the previous two is that each semicircle has a convex and a concave side, reminiscent of the passive zones considered by Cates \etal \cite{cates}. It was our expectation that the due to this asymmetry, there would be a net flow of particles and indeed the measurement of the mean velocity of the center of mass in the $x$ direction (\textit{i.e.} from left to right in Figure \ref{Boundaries}), $v_{\textrm{cm}}$, show that in the presence of asymmetric active patches, the system establishes a net flow. In Figure \ref{ParticleCannons}, we plotted $v_{\textrm{cm}}$ for a gas of active particles both in the absence and in the presence of interparticle interactions. The plot shows that in both cases, there is a flow and the error bars represent the variance of the observed velocity for $6$ independent simulation runs. In the inset of Figure \ref{ParticleCannons}, we plotted the progression of the center of mass in the $x$ and the $y$ direction as a function of time for a large activity difference and with interparticle interactions. One can see in the inset, that the flow in the $x$ is significantly larger than the typical fluctuation in the $y$ direction and the irregularity in the displacement time graph indicates that the flow is stochastic. The comparison between the ideal active gas and the active hard spheres show that there are two mechanisms at play. In the ideal gas case, the particles enter and exit the passive zone independently from each other and therefore the flow is the result of the interaction of the active particles and the geometry of the passive zone. When the interparticle interactions are switched on, the mean velocities are initially the same because the density in the passive zone is comparatively small, so exclusion effects are negligible. As the activity increases, so too does the crowding within the passive zone until, eventually, particles trying to enter the passive zone are kept at the boundary. Therefore, in addition to the geometrically induced flow of the ideal gas case, we observe active particles that are sliding along the convex half of the boundary with greater ease than along the concave half, as can be seen in the supplemental video V3, where a comparison is made between active particles with and without interparticle interactions using the same levels of activity. In the limit of very large activities, the particles in the passive zone crystallize and the ensuing domain acts like a solid boundary, similar to the passive tracers in an active bath that were studied by Caccutto \etal \cite{AngeloTracers}. In this final case, the analogy between the system with an activity difference and a temperature difference breaks down completely, since simulations of the latter did not exhibit any flow. 

\section{Conclusions and outlook}

Although a quantitative theory predicting the response of an activity difference is still pending, we have found that particles naturally tend to aggregate in regions of low activity. We examined the behavior of a system of strongly repulsive spheres with an activity difference, and found that they act qualitatively similar to a system with a nonuniform temperature profile, provided that the mobility differences are symmetric. Furthermore, it is possible to induce crystallization if the global density and activity difference are appropriately chosen and the shape of the crystalline domains is entirely determined by the geometry of the activity field. Mismatches in the curvature of the passive zones, tend to induce a flow in the system, unlike the purely diffusive case.

It is worth noting that this work considers one of the most primitive realizations of active matter. Although one might extend the model to include more complex potentials or include hydrodynamics, the most dramatic change is likely to occur if the shape of the active particles is altered. Active rods, for example, tend to form nematic and smectic phases, but it is unclear how a smectically ordered domain is affected by a bath of active rods swimming around it, especially if the shape of the domain is circular or elliptical. It is likely that a system of rods with an activity difference is a new class of liquid crystal. Finally, it is also expected that inserting particles of arbitrary composition and shape in this system will result in them being pushed into regions of low activity. 

This research was supported by the FWF under the project No. P24681-N20 and within the SFB ViCoM (F41). The authors would like to thank Luca Tubiana and Emanuele Locatelli for many useful discussions and suggestions.

\end{document}